\documentclass[aps,prl,preprint,superscriptaddress,a4paper]{revtex4}

\usepackage{graphicx,amssymb,amsmath}
\usepackage{dcolumn}
\usepackage{color}

\begin{document}
\bibliographystyle{revtex}

\title{Search for $\eta$-mesic nuclei in recoil-free transfer reaction}

\author{A.~Budzanowski}
\affiliation{Institute of Nuclear Physics, PAN, Krakow, Poland}

\author{A.~Chatterjee}
\affiliation{Nuclear Physics Division, BARC, Mumbai, India}

\author{P.~Hawranek}
\affiliation{Institute of Physics, Jagellonian University, Krakow,
Poland}

\author{R.~Jahn}
\affiliation{Helmholtz-Institut f\"{u}r Strahlen- und Kernphysik der
Universit\"{a}t Bonn, 53115 Bonn, Germany}

\author{B.~K.~Jain}
\affiliation{Mumbai University Mumbai, India}

\author{V.~Jha }
\email[present address:  Institut f\"{u}r Kernphysik, Forschungszentrum J\"{u}lich,
J\"{u}lich, Germany]{}
\affiliation{Nuclear Physics Division, BARC, Mumbai, India}

\author{S.~Kailas}
\affiliation{Nuclear Physics Division, BARC, Mumbai, India}

\author{K.~Kilian}
\affiliation{Institut f\"{u}r Kernphysik, Forschungszentrum J\"{u}lich,
J\"{u}lich, Germany}

\author{S.~Kliczewski}
\affiliation{Institute of Nuclear Physics, PAN, Krakow, Poland}

\author{Da.~Kirillov}
\affiliation{Institut f\"{u}r Kernphysik, Forschungszentrum J\"{u}lich,
J\"{u}lich, Germany}
\affiliation{Fachbereich Physik, Universit\"{a}t
Duisburg-Essen, Duisburg, Germany}

\author{Di.~Kirillov}
\affiliation{Laboratory for High Energies, JINR Dubna, Russia}

\author{D.~Kolev}
\affiliation{Physics Faculty, University of Sofia, Sofia, Bulgaria}

\author{M.~Kravcikova}
\affiliation{Technical University, Kosice, Kosice, Slovakia}

\author{M.~Lesiak}
\affiliation{Institute of Physics, Jagellonian
University, Krakow, Poland}
\affiliation{Institut f\"{u}r Kernphysik, Forschungszentrum J\"{u}lich,
J\"{u}lich, Germany}

\author{J.~Lieb}
\affiliation{Physics Department, George Mason University, Fairfax,
Virginia, USA}

\author{L.~C.~Liu}
\affiliation{Theoretical Division, Los Alamos National Laboratory,
Los Alamos, New Mexico 87545}

\author{H.~Machner}
\email[corresponding author: ]{h.machner@fz-juelich.de} \affiliation{Institut f\"{u}r
Kernphysik, Forschungszentrum J\"{u}lich, J\"{u}lich, Germany}
\affiliation{Fachbereich Physik, Universit\"{a}t Duisburg-Essen,
Duisburg, Germany}

\author{A.~Magiera}
\affiliation{Institute of Physics, Jagellonian University, Krakow,
Poland}

\author{R.~Maier}
\affiliation{Institut f\"{u}r Kernphysik, Forschungszentrum J\"{u}lich,
J\"{u}lich, Germany}

\author{G.~Martinska}
\affiliation{P.~J.~Safarik University, Kosice, Slovakia}

\author{S.~Nedev}
\affiliation{University of Chemical Technology and Metalurgy, Sofia,
Bulgaria}

\author{N.~Piskunov}
\affiliation{Laboratory for High Energies, JINR Dubna, Russia}

\author{D.~Proti\'c}
\affiliation{Institut f\"{u}r Kernphysik, Forschungszentrum J\"{u}lich,
J\"{u}lich, Germany}

\author{J.~Ritman}
\affiliation{Institut f\"{u}r Kernphysik, Forschungszentrum J\"{u}lich,
J\"{u}lich, Germany}

\author{P.~von Rossen}
\affiliation{Institut f\"{u}r Kernphysik, Forschungszentrum J\"{u}lich,
J\"{u}lich, Germany}

\author{B.~J.~Roy}
\affiliation{Nuclear Physics Division, BARC, Mumbai, India}

\author{P.~Shukla}
\affiliation{Nuclear Physics Division, BARC, Mumbai, India}

\author{I.~Sitnik}
\affiliation{Laboratory for High Energies, JINR Dubna, Russia}

\author{R.~Siudak}
\affiliation{Institute of Nuclear Physics, PAN, Krakow, Poland}

\author{R.~Tsenov}
\affiliation{Physics Faculty, University of Sofia, Sofia, Bulgaria}

\author{J.~Urban}
\affiliation{P.~J.~Safarik University, Kosice, Slovakia}

\author{G.~Vankova}
\affiliation{Institut f\"{u}r Kernphysik, Forschungszentrum J\"{u}lich,
J\"{u}lich, Germany}
\affiliation{Physics Faculty, University of Sofia,
Sofia, Bulgaria}

\collaboration{The COSY-GEM Collaboration}




\date{\today}

\begin{abstract}%
We have studied the reaction $p+{^{27}Al}\to {^3{He}}+p+\pi^-+X$ at
recoil-free kinematics. An $\eta$ meson possibly produced
 in this reaction  would be
thus almost at rest in the laboratory system and could therefore be
bound with high probability, if nuclear $\eta$ states exist. The
decay of such a state through the $N^*(1535)$ resonance would lead
to a proton-$\pi^-$ pair emitted in opposite directions. For these
conditions we find some indication of such a bound state. An upper
limit of $\approx$ 0.5 nb is found.
\end{abstract}


%

\maketitle

The study of $\Lambda$ and $\Sigma$ hypernuclei, which are nuclear
bound systems of short-lived hadrons, has proved to be a very useful
tool for gaining information about $\Lambda -N$ and $\Sigma -N$
interactions in nuclei. Also, studies of $\pi^-$- and $K^-$- atomic
levels have  provided useful information  about $\pi - N$ and $K -
N$ interactions in nuclei. However, there has never been, to the
best of our knowledge, an observation of a neutral pseudoscalar
meson bound strongly in the nucleus. Observation of such bound
states would open new possibilities in nuclear and particle physics
with respect to the structure of such nuclei, the $\eta NN^*$
coupling constant and the behavior of the S$_{11}$ nucleon resonance
in nuclei.

 In contrast to the pion-nucleon interaction,
the $\eta$-nucleon interaction at small momenta is attractive and
sufficiently strong. This attraction can be seen from the fact that
the $\eta$ threshold (1488~MeV) is situated just below the
$N^*(1535)$ resonance which couples strongly to the $\eta -N$
channel. Initial calculations by Bhalerao and Liu \cite{Bhalero85}
obtained attractive s-wave $\eta -N$ scattering lengths $a_{\eta
N}=(0.28 + 0.19 i)$~fm and $a_{\eta N} = (0.27 + 0.22 i )$~fm, using
the $\pi - N$ phase shifts calculated by Arndt and the CERN theory
group, respectively. With these phase shifts, Haider and Liu
\cite{Haider_Liu86} have shown that $\eta$ can be bound in nuclei
with  A $\ge$ 10. Other groups have also found similar results
\cite{Hayano99, Tsushima00, Garcia-Recio02}. Recent analyses of the
experimental data and different theoretical calculations predict a
range of values for the $\eta -N$ scattering length from 0.2~fm to
1.0~fm for the real part, and from 0.2~fm to 0.35~fm for the
imaginary part. The higher values for the real part of $a_{\eta N}$
have led to speculations that the $\eta$-bound state might be
possible even for lighter nuclei. An overview  of
 this topic is given in Ref.
\cite{Haider_Liu02}.

 There have been previous searches for the
proposed $\eta$-mesic nucleus. First experiments searching for
$\eta$-mesic nuclei at BNL \cite{Chrien88} and LAMPF \cite{Lieb88}
by using a missing-mass technique in the ($\pi^+, p$) reaction came
to negative or inconclusive results. Later
 it became clear that the peaks are not necessarily
narrow and that a better strategy of searching for $\eta$-nuclei is
required. Furthermore, the BNL experiment was in a region far from
the recoilless kinematics, in which the cross section is
substantially reduced \cite{Hirenzaki07}. \color[named]{Black} More
recently, the existence of $\eta$-mesic ${^3He}$ was claimed to have
been observed in the reaction $\gamma{^{3}He}\to \pi^0pX$ using the
photon beam at MAMI  \cite{Pfeiffer04}. It has, however, been
pointed out \cite{Hanhart05} that the data of Ref. \cite{Pfeiffer04}
does not permit an unambiguous determination of the existence of a
${^3He} \eta$-bound state. The suggestion that ${^3He} \eta$ is not
bound is also supported by the theoretical studies of Refs.
\cite{Sofianos97, Haider_Liu02}.

The present experiment makes use of the transfer reaction
\begin{equation}\label{equ:transfer_reaction}
p+{^AZ}\to {^3He}+{^{A-2}(Z-1)}\otimes\eta
\end{equation}
The ${^3}$He were measured at zero  degrees
 with the magnetic spectrograph Big Karl
\cite{Bojowald02} by ray tracing in the focal plane with two packs
of multi-wire chambers. This was followed by two hodoscope layers,
 separated by 4~m which provided an additional
time-of-flight measurement. The beam momentum ($p_\text{beam}=1745$
MeV/c) and the setting of the spectrograph were chosen such that for
binding energies in the range 0-20~MeV, the $\eta$ is produced
almost at rest. The ${^3}$He spectrum is expected to be dominated by
particles being emitted during the nuclear cascade process.
 In order to reduce this background a coincidence
was required among ${^3}$He and events produced through a second
step \color[named]{Black}
\begin{equation}\label{equ:step}
\eta+n\to \pi^-+p.
\end{equation}
Because the overall $\eta n$ system (or $N^*$) is almost at rest,
energy and momentum conservation require the two charged particles
to be emitted back-to-back to each other with energies of
$\approx$100~MeV for the proton and $\approx$348~MeV for the pion.
Such a clear pattern is smeared out by Fermi motion resulting in a
distribution around $\approx$150 $^\circ$ with a width of
40$^\circ$. For the measurement of these particles a dedicated
detector ENSTAR was built, details of which are described in Ref.\
\cite{Betigeri07}. Briefly, it consists of three cylindrical layers
of scintillating material surrounding the target. Each layer is
divided into long bars thus allowing a measurement of the azimuthal
emission angle.  The bars of the middle layer are further divided
along the length in order to measure the polar emission angle. While
the protons of interest are stopped in the middle layer of the
detector, pions pass through all layers giving only $\Delta E$
information.

 Although, some  calculations
predict ${^4}$He to be large enough to bind $\eta$ mesons,
Garcia-Recio et al. \cite{Garcia-Recio02} expect more medium mass
nuclei (A $\sim$ 24) to show stronger binding. On the other hand,
heavier nuclei will have broader states making them harder to detect
on a smooth background. Furthermore, the final nucleus should not
have too many excited states, which is the case for even-even
nuclei. The ideal target should thus be odd-odd, but such a nucleus
does not exist as a solid target so we were limited to an odd-even
system. As a compromise among these different factors
 we choose $^{27}$Al. The target thickness of 1~mm, corresponding to a resolution of 2~MeV, was chosen in order not
to spoil the natural width of the bound state. Two runs were
performed with different spectrometer momentum settings (p$_0$ = 859~MeV/c and 897~MeV/c).  An integrated luminosity of $0.50\pm 0.05$~pb$^{-1}$ was accumulated for each run. \color[named]{Black}

Prior to the experiment, the ENSTAR detector was calibrated as
described in Ref.\ \cite{Betigeri07}. Due to the high brilliance
proton beam the experiment was performed with minimal background
even though the $^3$He were measured in the forward direction.
\begin{figure}[h]
\begin{center}
\includegraphics[width=8 cm]{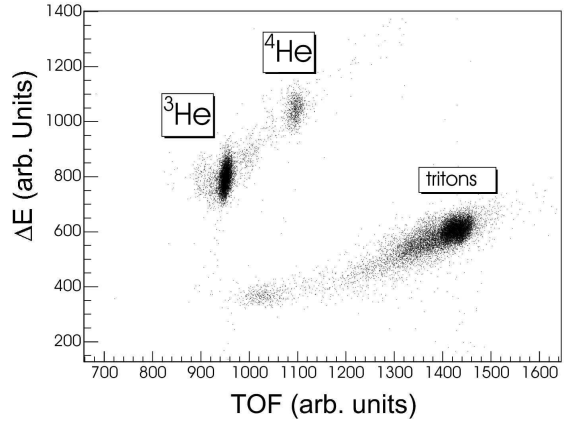}
\caption{Energy loss $\Delta E$ as function of the time-of-flight
(TOF) for particles in the focal plane area of the magnetic
spectrograph.} \label{Fig:PID}
\end{center}
\end{figure}

In Fig. \ref{Fig:PID} the energy loss in the first hodoscope layer
is shown as a function of the time-of-flight. Different particle
groups can be clearly identified. Beam particles do not enter the
focal plane because their charges differ by a factor of two. This
would not be the case in a deuteron induced reaction where break-up
protons would flood the focal plane detectors. The inclusive $^3$He
spectra are uniformly distributed when the data are corrected for
the acceptance of the spectrograph.  $^3$He within the angles
$\approx$ 3$^\circ$ in the vertical and $\approx$ 0.6$^\circ$ in the
horizontal direction were recorded by the spectrograph focal plane
detectors.

The coincidence required between the focal plane detectors and the
ENSTAR detector was achieved  by measuring the time between the
first hodoscope layer and one of the individual ENSTAR elements. A
peak-to-background ratio of 3.2:1 was obtained and the background
was subtracted.

\begin{figure}[h]
\begin{center}
\includegraphics[width=8 cm]{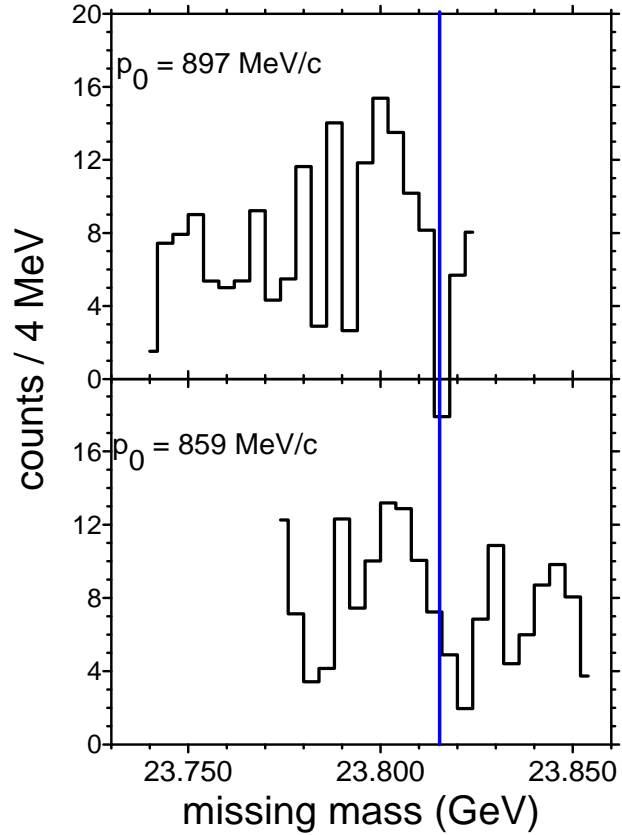}
\caption{(Color online) Missing mass spectra for two spectrograph
settings indicated in the figure. The counts were generated from
acceptance-corrected ${^3He}$ spectra measured in the magnetic
spectrograph with two charged particles detected in the ENSTAR
detector, which show the decay pattern of an $N^*$. The solid line
indicates zero excitation energy of a $^{25}Mg\otimes\eta$ system,
i. e. binding is to the left of the line.} \label{Fig:BE}
\end{center}
\end{figure}

For the beam momenta used, the selection of $^3$He means that the
residual system is at rest with an excitation energy of $\approx$
550~MeV. The only background which could give the same pattern as
the $N^*$ decay would be a deuteron, stopping in the middle layer in
association with a higher energy proton punching through all
detectors. A gate was therefore put on pions on a $\Delta E-E$
spectrum for events going through all layers. With
 such  geometrical selections
we obtain the missing-mass spectra for the two spectrograph settings
shown in Fig. \ref{Fig:BE}. The counts have been corrected for the
acceptances of the spectrograph for the two settings.  In order to
minimize systematical uncertainties in areas of small acceptance,
only the regions with acceptance 4$\%$ around the central momentum
value have been retained.  This eliminated data in less than $5\%$
of the missing mass range. Applying the cuts from ENSTAR
corresponding to $\eta+n\to \pi^-+p$ leads to reduction in yield by
a factor of $\sim 10^3$.
\begin{figure}[h]
\begin{center}
\includegraphics[width=0.39\textwidth]{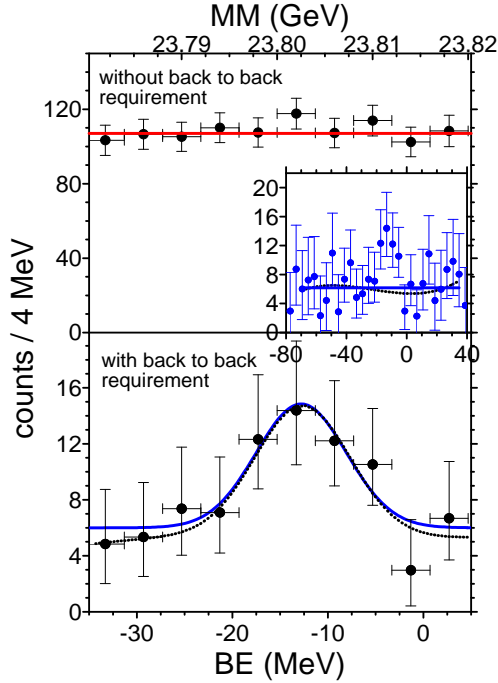}
\caption{(Color online) Data in the peak region as function of the
missing mass (upper abscissa) or the corresponding binding energy (lower abscissa). The upper panel shows the data
without the back-to-back correlation. The solid line in the upper
panel is a fitted constant to the whole data set. The solid curve
in the lower panel is a constant fitted as background and gaussian
on top of this background. The dotted curve is a fitted polynomial
as background and a gaussian on top of this background. The insert
shows the total data set for the back-to-back condition and two
different background fits to the data outside the peak region. The errors are asymmetric due to the underlying Poisson statistics.  }
\label{Fig:Binding-spectra}
\end{center}
\end{figure}
Positive values of the binding energy $BE$ correspond to the free or
unbound $\eta$ production. Due to the large width of the $N^*$
resonance of 100 to 200~MeV \cite{PDG}, the yield should rise with
phase space, while at the eta-mesic formation threshold it is expected  to be zero.
An indication of such a rise is somewhat seen in the data as
demonstrated by the fitted polynomial shown in the insert of Fig.\
\ref{Fig:Binding-spectra}, which gives a better $\chi^2$ value than
the one that is obtained by a fit with a
constant.

For both settings, there appears to be an enhancement from the threshold for
$^{25}Mg\otimes\eta$ which is -23.8145~GeV to $\approx -23.79$~GeV.
One may attribute all the counts to background. However, background
should be randomly distributed and hence it is very unlikely for two
different measurements to show the same structure. We, therefore,
conclude, the structures could be from the bound $\eta$.

In order to elucidate this point  further  we discuss the spectra in
more detail. In Fig.\ \ref{Fig:Binding-spectra} we show the binding
energy spectra combined for the two settings. Since for both
settings the same luminosity was acquired, the weighted arithmetic
mean was used in overlap region.  The figure shows the data without
the strong back-to-back correlation requirement (upper panel) and
with the requirement (lower panel). The unconstrained data do not
show any structure and can be well described by a constant.
 For the data in the lower spectrum, the $N^*$
decay pattern is required. The  counts are
typically lower by an order of magnitude than that in the
unconstrained case. Although, the $N^*(1535)$ can also decay with
two-pion emission, this branch is small compared to the $p\pi^-$
channel \cite{PDG}.

The data show an enhancement around  $BE\approx
-13$~MeV. The significance of this structure is
extracted according to the two methods given in Refs. \cite{FST_statistics} and \cite{Belle} respectively. At first, we test the hypothesis of peak
structure being fluctuation of background, i.e. the origin of
the background is taken to be independent of the signal.
The background outside the peak region, for simplicity approximated by a constant, was found to be 5.8$\pm$0.64. The significance \cite{FST_statistics} is then given by $(N-BG)/\sqrt{BG+\sigma_{BG}}$
where $N$ is the total counts in the region of interest, $BG$ is the
total background in this region as determined from the fit to the
outside region and $\sigma_{BG}$ is error in the estimation of
background value as taken from the fit. This yields a value of
significance which is 5.3$\sigma$. Here we have assumed Gaussian errors. For the assumption of Poisson errors with asymmetric error bars (see Fig. \ref{Fig:Binding-spectra}) the background is 6.2$\pm$1.0. This larger value is typical for Poisson distribution and hence the significance reduces to 4.9$\sigma$. Finally a Gaussian on top of the background was fitted to the whole data set. This yielded for the case of Poisson statistics 6.4$\pm$0.96 for the background, 8.3$\pm$3.6 for the amplitude, -12.0$\pm$2.2~MeV for the centroid and 4.7$\pm$1.7~MeV for the width. The corresponding curves are shown in Fig. \ref{Fig:Binding-spectra}. Also a third order polynomial was fitted to the data. The resulting curves are also shown in Fig. \ref{Fig:Binding-spectra}.

In the second method, the statistical significance is extracted by assuming the background events as well as the peak events on top of the background being Poisson distributed. Again a constant background and a Gaussian was assumed. A fit was performed using the maximum likelihood method.
The significance is then defined as, $\sqrt{-2  \Delta ln L}$. Here,
$\Delta ln L$ is the difference in the values of the logarithm
likelihood function with signal fixed to zero and at the best fit
value. In this way, we obtain a value of 6.20$\sigma$ for the
significance, assuming a simultaneous determination of amplitude,
centroid and width of the signal. The fit gives for the linear
background $6.38\pm 0.53$ together with the values, for the signal
amplitude $8.55\pm 3.05$, for the centroid $-13.13\pm 1.64$~MeV and
for the width $4.35\pm 1.27$~MeV corresponding to a FWHM of
10.22$\pm$2.98~MeV. These results compare favorably with those from the first method. We, therefore, consider the present experimental results to provide
a strong hint of a nuclear $\eta$ bound state.

This allows us to give an upper bound for the cross section. With an
estimated efficiency due to detector geometry and analysis
selections of 0.70 $\pm$ 0.07 we find $\sigma$ = 0.152 $\pm$ 0.054
(stat) $\pm$ 0.021 (syst)~nb. If this ``structure" corresponds to a
bound $\eta$ decaying via reaction (Eqn.\ \ref{equ:step}), the
 cross section  would be 0.46
$\pm$ 0.16 (stat) $\pm$ 0.06 (syst)~nb, assuming an isospin
branching ratio of 1/3. This cross section value
 can be compared with the ``elementary"
$pd\to{^3He}\eta$ reaction which, for the present beam energy has a
cross section integrated over the spectrograph acceptance of
$\approx 39 \mu$~b \cite{Berthet85, Banaigs73}.

In summary, we have measured the reaction $p+^{27}Al\to
{^3He}+\pi^-+p+X$. The ${^3}$He ions, which were detected at zero
degrees with a magnetic spectrograph, carried the beam momentum
(recoilless kinematics).   The remaining system has the mass $m(^{25}$Al$)+m(\eta)+BE$ with BE the binding energy. The $\pi^-+p$ system, measured with the
ENSTAR detector, decays almost back-to-back with energies
corresponding to an $N^*(1535)$ at rest. The most probable scenario for the $\eta$ decay is through forming an in-medium $N^*$ resonance which decays into $p$ and $\pi^-$. In this case, the remaining system is $m(^{24}\text{Al}+m(N^*(1535)_\text{in medium})$. Although instead of an $\eta$ a pion could be produced in the intermediate step forming another $N^*$ nearby, which could lead to the back-to-back $\pi^-p$ events. However, simple kinematical calculations show that this would require the momentum of a target nucleon to at least 210~MeV/c for which the probability is very low.  In two spectra, taken at
different spectrograph settings, an enhancement was found for
negative binding energies close to the free production threshold.
This is exactly what is expected from the bound $\eta N$ system. The
enhancement may not be purely due to binding in the ground state
only but also to an excited $^{25}Mg$ state. However, this requires
pick up of more deeply lying nucleons which may be less likely than
pick up of the least-bound nucleons. Binding energy spectra without
strong  $N^*$ constraint do not show the enhancement.

\section{Acknowledgements} Discussions with B.~Kamys and C.~Wilkin
are gratefully acknowledged. We appreciate the support received from
the European community research infrastructure activity under the
FP6 "Structuring the European Research Area" program under the
contract no. RII3-CT-2004-506078, from the Indo-German bilateral
agreement, from the Research Center J\"{u}lich (FFE), and from GAS
Slovakia (1/4010/07).



\end{document}